\documentclass{article}
\usepackage{optidef}
\PassOptionsToPackage{numbers}{natbib}
\usepackage{graphicx}
\usepackage[final]{neurips_2022_ml4ps}

\usepackage[utf8]{inputenc} 
\usepackage[T1]{fontenc}    
\usepackage{hyperref}       
\usepackage{url}            
\usepackage{booktabs}       
\usepackage{amsfonts}       
\usepackage{nicefrac}       
\usepackage{microtype}      
\usepackage{xcolor}         
\usepackage[normalem]{ulem}

\title{Deep-pretrained-FWI: combining supervised learning with physics-informed neural network}

\author{Ana Paula O.~Muller\\
Centro Brasileiro de Pesquisas Físicas (CBPF)\\
Rua Xavier Sigaud, 150, Urca, Rio de Janeiro, Brazil. \\
Petróleo Brasileiro S.A. (PETROBRAS),\\
Edifício Senado, Av. Henrique Valadares, 28,Centro, Rio de Janeiro, Brazil.\\
\And
Clecio R.~Bom\\
Centro Brasileiro de Pesquisas Físicas (CBPF)\\
Rua Xavier Sigaud, 150, Urca, Rio de Janeiro, Brazil.\\
Centro Federal de Educação Tecnológica Celso Suckow da Fonseca (CEFET-RJ),\\
Av. Maracanã, 229,Maracanã, Rio de Janeiro, Brazil.
\And
Jessé C.~Costa\\
Universidade Federal do Pará, Belém, PA, Brazil.\\
National Institute of Petroleum Geophysics (INCT-GP)
\And
Matheus Klatt, Elisângela L.~Faria, Marcelo P.~ de Albuquerque, Márcio P.~ de Albuquerque \\
Centro Brasileiro de Pesquisas Físicas (CBPF)\\
Rua Xavier Sigaud, 150, Urca, Rio de Janeiro, Brazil.\\
}

\begin{document}

\maketitle

\begin{abstract}
An accurate velocity model is essential to make a good seismic image. Conventional methods to perform Velocity Model Building (VMB) tasks rely on inverse methods, which, despite being widely used, are ill-posed problems that require intense and specialized human supervision. Convolutional Neural Networks (CNN) have been extensively investigated as an alternative to solve the VMB task. Two main approaches were investigated in the literature: supervised training and Physics-Informed Neural Networks (PINN). Supervised training presents some generalization issues since structures, and velocity ranges must be similar in training and test set. 
Some works integrated Full-waveform Inversion (FWI) with CNN, defining the problem of VMB in the PINN framework. In this case, the CNN stabilizes the inversion, acting like a regularizer and avoiding local minima-related problems and, in some cases, sparing an initial velocity model.
Our approach combines supervised and physics-informed neural networks by using transfer learning to start the inversion. The pre-trained CNN is obtained using a supervised approach based on training with a reduced and simple data set to capture the main velocity trend at the initial FWI iterations. We show that transfer learning reduces the uncertainties of the process, accelerates model convergence, and improves the final scores of the iterative process.

\end{abstract}

\section{Introduction}
The construction of an accurate seismic image is critical for the achievement of a good velocity model. Full-waveform inversion (FWI) aims to recover the velocity model from the acquired shots, using an iterative inverse method that updates the initial guess for the velocity model subsided by the wave-propagation equation \cite{tarantola,pratt}. In theory, FWI can recover the velocity model with high accuracy and details; however easily achieves a local minimum, leading the model solution in the wrong direction. 

Recently machine learning has attracted great interest in solving the VMB problem \cite{reviewMLvel}. There are a diversity of proposed solutions on the theme which rely on supervised training \cite{SALTTariq,klatt2022,muller20221,ARAYAPOLO1,ARAYAPOLO2,YANGBASE,shucai20,attentiongate}. The common approach is to train the network using synthetic data since there is no means to know exactly the true velocity model related to the real seismic data, which leads to generalization and domain adaptation problems \cite{intsalt}, for instance, wavelet match with the acquired data \cite[]{paper4}, and dependency on the model velocity features on training data \cite{Shragge}, to name a few. 

Physics-informed neural networks (PINN) have emerged as an alternative to overcome the issue of the large amount of data required by supervised learning and have been adopted in geophysics to solve a diversity of problems \cite{PINNGeo1,PINNGeo2,PINNGeo3,COLOMBO2021,magnetotel2022}. Some works solved the VMB problem by integrating CNN architectures to FWI algorithms \cite{FWIHeWang,FWIMcMechan01,FWIBiondo,FWIMcMechan02}, where CNN regularizes the inversion but still requires an initial model. Recently, an approach suppressed the need for an initial model \cite{fwi_tle} by inputting the shots acquisition on an autoencoder, capturing the prior information from shots, and making the DL-FWI converge to a solution, despite obtaining almost random velocity models at the initial iterations. One point to consider is that this approach's convergence is very slow, with at least $1000$ iteration steps until reaching a reasonable velocity model.

To mitigate the observed slow convergence, we used a U-Net-based architecture, which requires few parameters and is commonly used to solve the VMB problem \cite{YANGBASE,muller20222}. We also proposed transfer learning to initialize the network's weights closer to the final solution. Instead of randomly initializing the CNN for DL-FWI, our transfer learning consists of loading the weights of pre-trained CNN specialized in predicting the velocity model from the acquisition shots in a supervised framework. Transfer learning is successfully applied in seismic problems \cite{TFGeo1,TFGeo2,TFGeo3,TFGeo4,TFGeo5,TFGeo6,TFGeo7}.  With this approach, we combine the strategy of the early works that are pure data-driven \cite{ARAYAPOLO2,YANGBASE,shucai20} with physics-informed neural networks. 

\section{Connecting FWI with PINN}

FWI is based on the idea that if we know the correct velocity model, we can simulate the acquired seismic shots $d_{obs}$. For the acoustic case, the process starts by applying the acoustic wave equation propagator $\mathbf{F}$ over an estimated velocity model $\hat{m}$, obtaining a simulated data $d_{sim}$. The misfit between $d_{obs}$ and $d_{sim}$ gives us a measure of how much our initial guess to the velocity model $\hat{m}$ is far from the real model $m$. The FWI solves an optimization problem by minimizing the misfit function with the adjoint state method \cite{adjoint}, which calculates the gradients to update the velocity model $\hat{m}$. Then, a new iteration starts, and the updated velocity model is again availed with FWI. The process repeats until a satisfactory velocity model is found. FWI is an ill-conditioned inversion, highly dependent on the initial model \cite{VIRIEUXFWI}.

One possible way of using PINN to obtain the velocity model from the seismic data is by training a CNN to generate a representation of the initial velocity model for the FWI \cite{FWIMcMechan01, FWIHeWang, FWIBiondo}. Usually, the input of such CNN is a random vector or a coordinate vector, and the main goal with such velocity model representation is to regularize the inversion process. Recently, a methodology eliminated the initial model's dependency \cite{fwi_tle} by combining the regularization power of CNN with the idea of inputting the shots into the CNN to determine the velocity model. For convenience, we will call this approach DL-FWI. The information flow of DL-FWI is summarized in Figure \ref{fig:DL-FWI}(a). Since the CNN weights are randomly initialized by default, the models obtained with the DL-FWI at the initial iterations present irregular structures and are very far from the real ones. The process slowly converges to a solution close to the real model, which would be impossible without coupling the FWI algorithm to the CNN. However, the slow convergence is unfeasible when considering extrapolating the method to real cases.

\subsection{Implementation details}

\begin{figure}
  \centering
\includegraphics[width=0.65\textwidth]{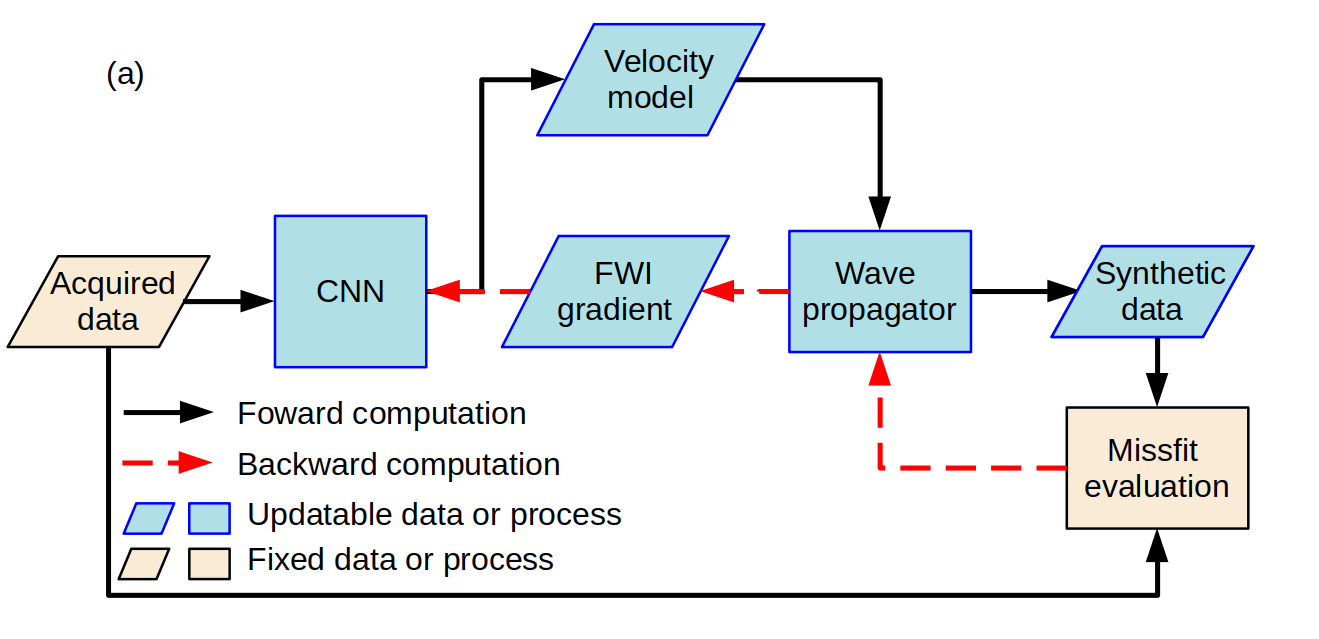}
\includegraphics[width=0.32\textwidth]{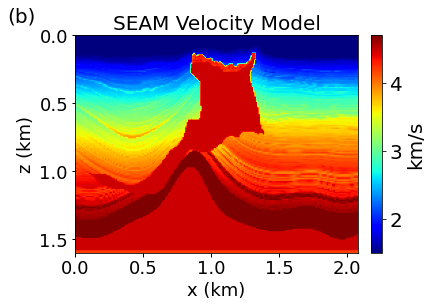}
  \caption{Figure (a) summarizes the DL-FWI information flow. Figure (b) shows the SEAM velocity model.}  
  \label{fig:DL-FWI}
\end{figure}

We performed our experiments using one GPU V100, with the \textit{Pytorch} and \textit{Deepwave} packages \cite{pytorch,DeepWave}. We show the results for one benchmark model, the SEG Advanced Modeling Corporation (SEAM) Phase 1 model \cite{seam}, which presents complex structures and high-velocity contrasts. As shown in Figure \ref{fig:DL-FWI}(b), the velocity values were kept equal to the original benchmark models, and the dimensions were re-scaled to be equal to 2.08kmx1.6km, with a grid space equal to 10m. The seismic acquisition was configured with receivers in a fixed spread geometry, with one receiver at each model position. The shots are made every 100 meters, starting at 40 meters from the model border, with a total of twenty shots, each shot act like a channel of the input image. The source wavelet is a Ricker, with a peak frequency of 8Hz. The data is registered for 3.2 seconds, using a sampling rate of 0.001s. 

The problem to be solved here has unbalanced vertical dimensions since we are inputting the time samples of the registered shots and outputting the vertical samples of the velocity model. In our work, we used a U-Net due to the previous success of such architecture in solving seismic analysis problems \cite{faults_U,muller20221,muller20222,SALTUNET,DENOISE}, and to reduce the number of trainable parameters when compared with the Autoencoder used in the first DL-FWI implementation \cite{fwi_tle}.
The U-Net receives a [208,160,20] volume as input, with the input channels corresponding to the number of shots simulated, and produces a [208,160,1] probability map between zero and one as output, which is then rescaled to the expected range of the velocity model. Our network starts with 36 feature maps in the first stage of the encoder and presents four stages in the encoder section, which gradually downsamples the data. The decoder section has four stages and gradually upsamples the feature map back to the original spatial dimensions.
The U-Net structure requires that the encoder and decoder branches present the same spatial dimensions, thus we tested a simple modification, which is to downsample the shots to get 160 vertical samples as input to the CNN. The shots used to evaluate the FWI misfit are kept with the original time sampling. To reach the best performance for the U-Net, it was necessary to modify the original architecture. The seismic shots imprinted some diagonal artifacts in the shallow portions of the model predicted by the U-Net, which correlate with the shallow reflections on the shot data. Then, we removed the first shortcut connection to remove these artifacts. 

\begin{figure}
\begin{minipage}[c]{0.48\textwidth}
   \includegraphics[height=0.37\textwidth]{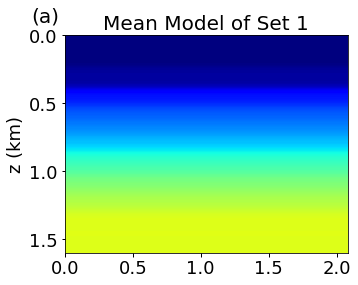}
  \includegraphics[height=0.37\textwidth]{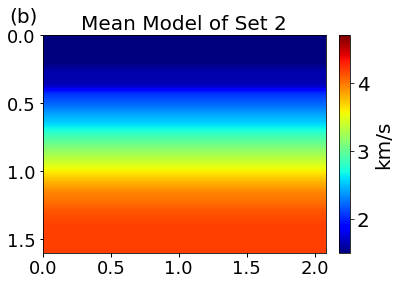}\\
  \includegraphics[height=0.40\textwidth]{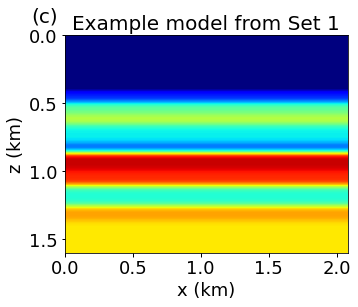}
  \includegraphics[height=0.40\textwidth]{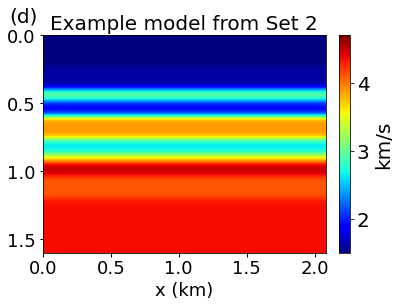}
\end{minipage}
\begin{minipage}[c]{0.50\textwidth}
\includegraphics[width=0.95\textwidth]{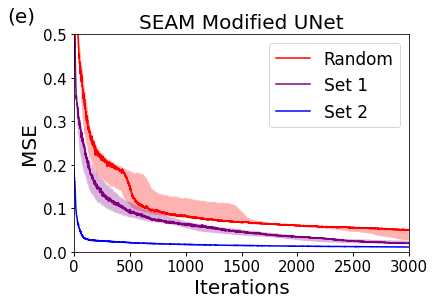}
\end{minipage}
  \caption{Figures (a) and (b) show the mean of the 240 velocity models of each Set used for pre-training the U-Net. Figures (c) and (d) show one velocity model sample from each Set. Figure (e) shows the results for each different initialization of the U-Net weights. The shaded area around the curves accounts for the stochastic effects of the method with the limits of the shaded area defined by the standard deviation of the observed MSE.}
  \label{fig:set1xset2}
\end{figure}

Training data sets were constructed using horizontal layers. We created two sets whose mean velocities are plotted in Figures \ref{fig:set1xset2}(a) and (b). In Figures \ref{fig:set1xset2}(c) and (d), we show one velocity model example of each Set. The models are simple and completely different from the SEAM model that we aim to solve with DL-FWI (Fig.~\ref{fig:DL-FWI}(b)). These simple training models were made intentionally to test the hypothesis that a weak training process with simple models is sufficient to speed up the convergence of DL-FWI. Each set presents 240 models, with ten horizontal layers of random thickness. The 240 were separated into 200 samples for training and 40 for validation. The velocity of each layer is defined randomly, in a crescent trend, but in a process that allows some layers of low velocity and high velocity. The only imposed difference in the flow that generated the two sets is the maximum velocity of the deepest layer, equal to 3500m/s for Set 1 and 4500m/s for Set 2. 

We trained the U-Net in a supervised way, using as input the seismic shots and evaluating an MSE loss function that compares the velocity model obtained from network prediction and the real velocity model. The U-Net was trained over 100 epochs with the Adam \cite[]{ADAM} variation of the stochastic gradient descent algorithm, with a base learning rate of $1e^{-3}$. Our transfer learning process consists of loading the weights of this pre-trained U-Net to initialize the DL-FWI process.

\begin{figure}
  \centering
  \includegraphics[width=\textwidth]{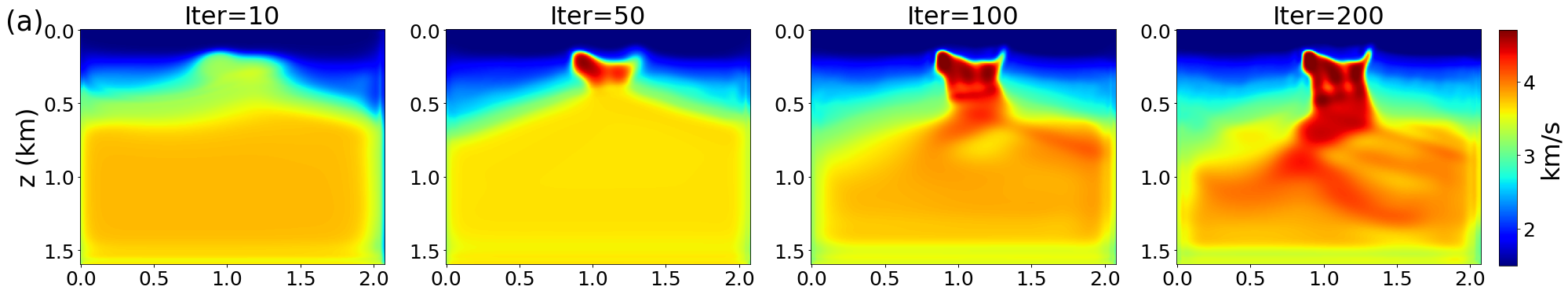}
   \includegraphics[width=\textwidth]{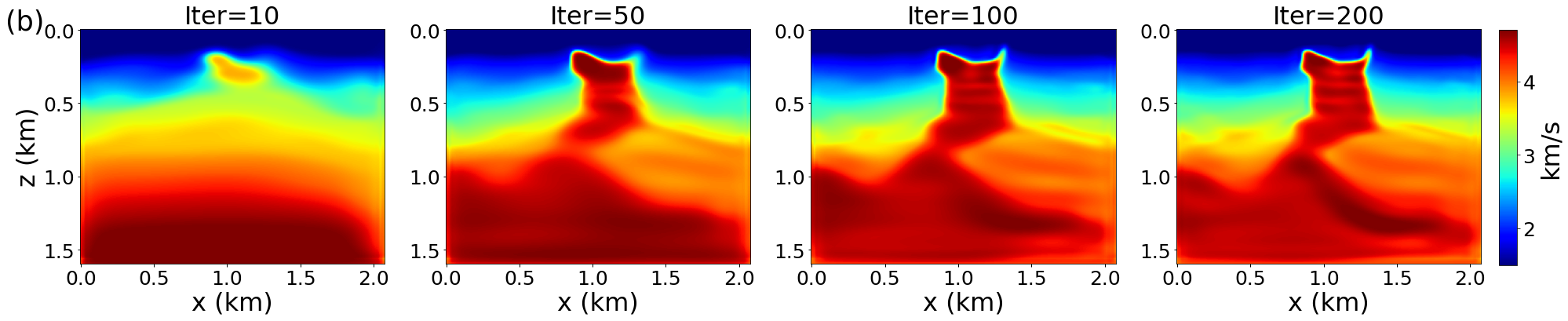} 
  \caption{Figure (a) shows the velocity along iterations for the U-Net with the random initialization, and figure (b) for the U-Net with transfer learning from Set 2.} 
  \label{fig:seam_evol}
\end{figure}

\section{Results}

The U-Net weights are randomly initialized when not using the transfer learning approach. Besides, the FWI misfit calculus is performed over a random mini-batch of shots; both features add stochasticity to the iterative DL-FWI process. Then, one adequate way to evaluate the results is to make a pool of independent inversion processes and statistically evaluate the velocity model evolution. We ran the iterative DL-FWI twelve times for each tested approach, showing the loss evolution curves as the median measure with a shadowed area corresponding to two standard deviations (Fig.~\ref{fig:set1xset2}(e)).

In Figure \ref{fig:set1xset2}(e), we show the evolution of MSE when using the modified U-Net with and without transfer learning. When not using transfer learning, i.e., the Random curve, a larger standard deviation during convergence is observed. The random approach generates the most unstable results, with some runs generating accurate velocity models and others with poor predictions. This behavior is not desirable in exploration seismic when the goal is always to reduce uncertainty. Concerning the transfer learning approach, represented by Set 1 and Set 2 curves, it is possible to observe that besides reducing the instabilities, the final MSE is generally smaller and decays faster when compared with the Random initialization. We can also correlate the best performance achieved with Set 2 with the similarity between the mean of Set 2 and the velocity trend of the true model.

To make clear the benefits of transfer learning, we show the model evolution along 200 iterations. In Figure \ref{fig:seam_evol}, it is possible to observe that the random initialized U-Net (Fig.~\ref{fig:seam_evol}(a)) is too far from the real model (Fig.~\ref{fig:DL-FWI}(b)) and has only captured the shallow salt shape. Figure \ref{fig:seam_evol}(b) shows the fast convergence for the pre-trained U-Net, where the low-frequency features of the model were completely resolved with 50 iterations, and the residual improvements of the subsequent iterations were related to sharpening the salt and other interfaces.

\section{Conclusions}

Besides the accuracy and stability of the results, another important commitment in seismic developments is the feasibility of the proposed implementation, particularly concerning the computational cost of the process. 
We probe simple structural models to define a lighter process of supervised training. The physics-driven part of the flow then defines the fine-tuning of the structural features of the velocity model. We evaluated the results of the proposed methodology with one benchmark velocity model with highly complex structures. The results showed that the random initialization of the CNN, i.e. without transfer learning, generates velocity models at the initial iterations that are too far from the right solution, which makes the convergence of the iterative process slow and adds a high degree of uncertainty to the process. When testing the transfer learning approach, we observed that the training with simple plane parallel models was enough to improve results and get a fast convergence. The method proposed in this work, as other DL methods combined with FWI \cite{fwi_tle,FWIBiondo}, are not yet feasible to be applied over real seismic data, shots with fixed spread geometry, except ocean bottom data, are not common in marine seismic acquisition, besides even if available, their size would exceed the memory resources. The extension to real data must investigate different data arrangements and 3D geometries.

\begin{ack}
APOM thanks Petrobras for sponsoring her postdoctoral research and for the permission to publish this work. 
JCC acknowledges the CNPq financial support through the INCT-GP and the grant 312078/2018-8 and Petrobras. CRB acknowledges the financial support from CNPq (316072/2021-4) and FAPERJ (grants 201.456/2022 and 210.330/2022). The authors acknowledge the LITCOMP/COTEC/CBPF multi-GPU development team for supporting the Artificial Intelligence infrastructure and Sci-Mind’s High-Performance multi-GPU system and to the SENAI CIMATEC Supercomputing Center for Industrial Innovation for the cooperation, supply, and operation of computing facilities.

\end{ack}
\bibliographystyle{unsrtnat}
\bibliography{ref}

\end{document}